\documentclass[12pt]{iopart}

\usepackage{graphicx}
\usepackage{dcolumn}
\usepackage{bm}

\begin{document}

\title{Doppler-free laser spectroscopy of buffer gas cooled molecular radicals}

\author{S. M. Skoff, R. J. Hendricks, C. D. J. Sinclair, M. R. Tarbutt, J. J. Hudson, D. M. Segal, B. E. Sauer and E. A. Hinds}

\address{Centre for Cold Matter, Blackett Laboratory, Imperial College London, Prince Consort Road, London, SW7 2AZ. United Kingdom.}

\begin{abstract}
We demonstrate Doppler-free saturated absorption spectroscopy of cold molecular radicals formed by laser ablation inside a cryogenic buffer gas cell. By lowering the temperature, congested regions of the spectrum can be simplified, and by using different temperatures for different regions of the spectrum a wide range of rotational states can be studied optimally. We use the technique to study the optical spectrum of YbF radicals with a resolution of 30\,MHz, measuring the magnetic hyperfine parameters of the electronic ground state. The method is suitable for high resolution spectroscopy of a great variety of molecules at controlled temperature and pressure, and is particularly well-suited to those that are difficult to produce in the gas phase.
\end{abstract}

\maketitle


\section{Introduction}
\label{sec:introduction}
Dense samples of atoms and molecules can be produced at low temperatures using the buffer gas method \cite{Doyle95}. The atoms or molecules of interest are introduced into a cryogenic cell containing cold helium gas and are cooled by collisions with the helium. Often, it is desirable to remove the molecules from the cell once they have cooled, separating them from the helium and delivering them to an experiment. This can be done by guiding away the fraction that flows through a hole in the cell wall using either magnetic or electric guides \cite{Doyle05, Patterson07, Rempe09}. Other experiments can be done directly inside the cell. In particular, the atoms or molecules can be trapped magnetically once they are cold enough \cite{Kim97, Weinstein(1)98, Weinstein(2)98, Campbell07}, and measurements of collision cross-sections can be made in the cell e.g. \cite{Campbell07, Ball98, Maussang05, Lu09}. The simultaneous magnetic trapping of atoms and molecules has also been demonstrated \cite{Hummon08}, and the combination of laser ablation and buffer gas cooling has been used to create a coherent optical medium with very high optical depth \cite{Hong09}. The cold buffer gas cell can also be used as a versatile spectroscopic tool, and here it offers enormous potential for the elucidation of molecular structure and chemical bonding and the measurement of molecular parameters. A huge variety of molecules can be produced in the cell by laser ablation of suitable precursors, including radicals and refractory compounds that are otherwise very difficult to produce and study in the gas phase. In the earliest work on buffer gas cooling, millimeter and submillimeter spectroscopic techniques were applied to the cooled molecules \cite{Messer84, Willey(1)88, Willey(2)88}, taking advantage of the narrow linewidths, the larger signals, and the simplification of the spectra that result from the cooling. Later, laser spectroscopy, using absorption or fluorescence detection, was applied to several buffer-gas-cooled molecular species, e.g. VO and PbO \cite{Weinstein(3)98, Egorov01}, but with Doppler-limited resolution.

To resolve finer details in the spectrum and measure transition frequencies with higher precision, Doppler-free laser spectroscopy~\cite{Demtroder02} is needed. There are several techniques for obtaining spectra free from Doppler broadening, including saturated absorption spectroscopy~\cite{Borde70,Haensch71}, two-photon spectroscopy~\cite{Cagnac73}, polarisation spectroscopy~\cite{Wieman76, Stert78} and velocity-selective optical pumping~\cite{Pinard79}.  The crucial feature of all of these is the presence of pump and probe laser beams that counter-propagate through the sample, leading to a narrow feature in the spectrum when a velocity class is simultaneously Doppler-shifted into resonance with both beams. For example, when the counter-propagating beams have the same frequency, they interact with different velocity classes at all frequencies other than the unshifted transition frequency. These Doppler-free techniques have been applied to a huge number of atomic and molecular systems, including radicals produced in flames e.g.~\cite{Reichardt00}, gas-phase reaction cells e.g.~\cite{Ernst83} and discharge tubes e.g.~\cite{Barbieri90}. Because the molecules formed in these sources are hot, the population is typically distributed over a large number of high rotational states, often resulting in small signals and a congested spectrum that can be difficult to interpret. In this paper, we demonstrate saturated absorption spectroscopy of radicals produced by laser ablation and cooled to low temperature in a cryogenic buffer gas cell. The method is a simple and versatile way of obtaining high-resolution optical spectra of species that are otherwise difficult to produce in the gas-phase. The temperature and pressure can easily be controlled over a very wide range. Using laser ablation of a target inside the cell, we produce a high density of cold YbF, a molecule that is of considerable interest because of its high sensitivity to the electric dipole moment of the electron~\cite{Hudson02}. The technique is suitable for studying the spectra of a wide variety of other molecules and is complementary to supersonic beam techniques~\cite{Scoles88} which similarly provide cooling of the translational and rotational temperatures, allow radicals to be introduced, and offer high spectral resolution and sensitivity.


\section{Experimental setup}
\label{sec:setup}

\begin{figure}
\begin{centering}
\includegraphics[width=0.7\textwidth]{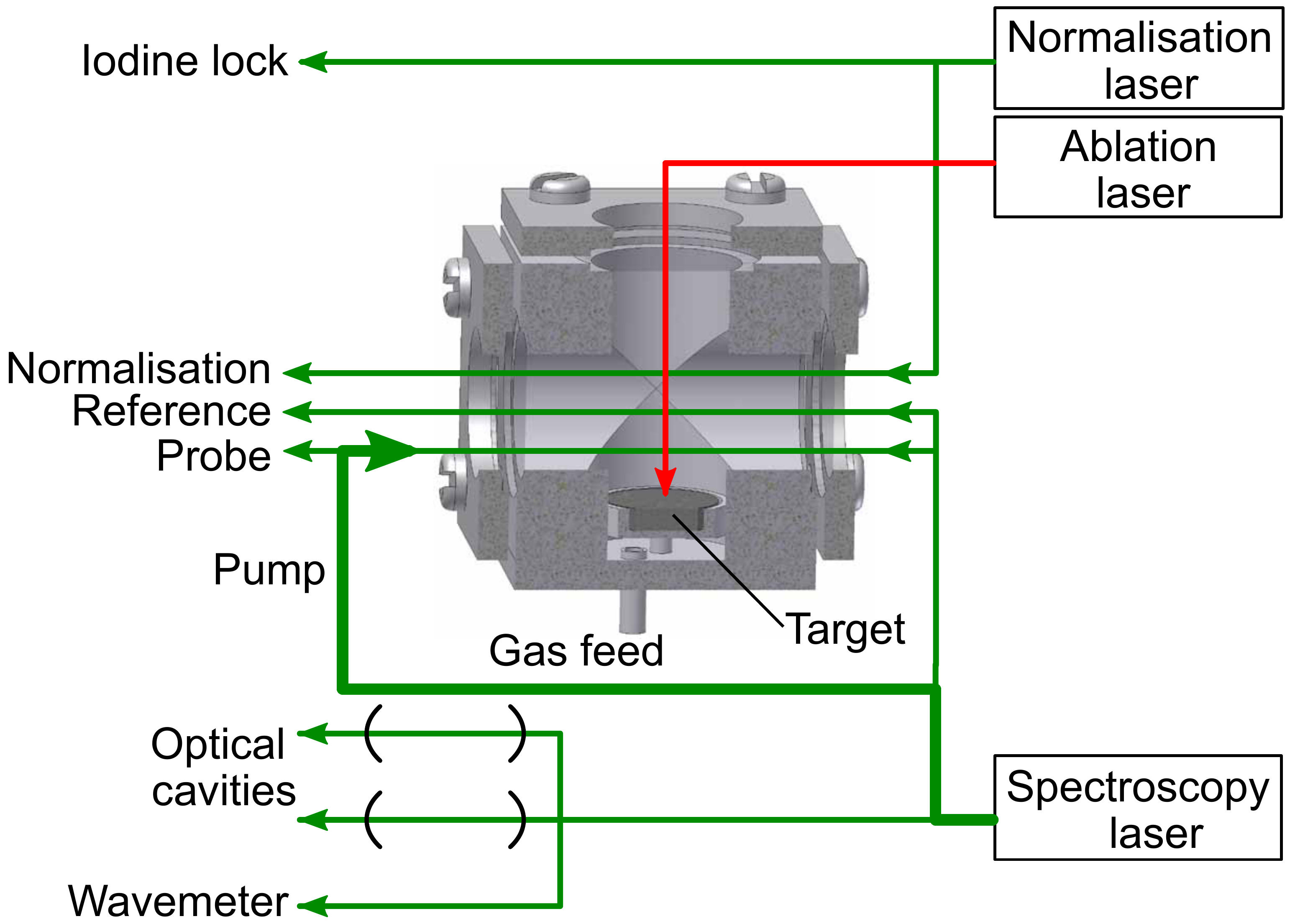}
\caption{\label{fig:spectroscopysetup}Schematic of the experimental setup}
\end{centering}
\end{figure}

Figure \ref{fig:spectroscopysetup} illustrates the experimental setup. An aluminium cell is mounted on the cold plate of a small liquid helium cryostat (Infrared Laboratories, model HD3) which is maintained at a pressure of approximately 10$^{-6}$ mbar by a turbomolecular pump. The cell is a cube of side 40\,mm with 20\,mm diameter windows in three of the faces and an ablation target mounted in a holder opposite one of these windows. All joints are made using vacuum-tight indium seals. A radiation shield heat sunk to the liquid nitrogen dewar prevents room temperature radiation from reaching the buffer gas cell. Indium tin oxide coated windows in this shield provide optical access at visible wavelengths, whilst reflecting thermal radiation. The temperature of the cell can be controlled with a heater on the cold plate, and by the choice of cryogen (helium or nitrogen).

Helium gas of 99.996\% purity enters the cell through a small hole located next to the ablation target. The gas is delivered through a tube which has an internal diameter of 2.5\,mm and is made of thin-walled stainless steel over most of its length to minimize heat conduction. To ensure that the gas is fully cooled when it enters the cell, the part of the tube near the cell is made of copper and is thermally connected to the cold plate of the cryostat. The pressure is measured at the room temperature end of the tube where the gas enters through a leak valve. By altering this, and a second valve connected to a rotary vane pump, the buffer gas pressure can be varied. In a subsidiary experiment, we connected a second pressure gauge directly to the cell and verified that when it is at room temperature the cell pressure is the same as the pressure at the usual measuring point. When the cell is cold, there is a thermomolecular pressure drop between the two ends of the gas delivery tube, which we determine using the empirical formula given in \cite{Roberts56}. The corrected cell pressure is then converted to a helium density using the measured cell temperature.

To create YbF molecules in the cell, a vacuum-hot-pressed target consisting of 70\% Yb and 30\% AlF$_3$ by mass (Sophisticated Alloys), is ablated by light from a 1064\,nm Nd:YAG laser delivering pulses with a fluence of approximately 1\,J/cm$^2$ (approximately 8\,ns duration, 30\,mJ energy and 2\,mm beam diameter at the target).

For the spectroscopy, light from a cw dye laser (Spectra Physics model 380, Rhodamine 560 dye) is split into three 1\,mm diameter beams --- probe and reference beams, each with an intensity in the cell of approximately 15\,mW/cm$^{2}$, and a counter-propagating pump beam with an intensity of about 300\,mW/cm$^{2}$ which overlaps the probe beam inside the cell. Two photodiodes monitor the absorption of the probe and reference beams. A small fraction of the laser output is coupled into a wavemeter which measures the frequency with an accuracy of 600\,MHz, and into two low finesse confocal optical cavities with free spectral ranges of 150\,MHz and 750\,MHz respectively. The transmission of these cavities provides a relative frequency scale with a precision better than 10\,MHz.

The yield of YbF molecules falls after multiple ablation pulses, due to degradation of the target. In order to obtain a reliable measure of the relative molecular yield as the spectroscopy laser is scanned, light from a second dye laser (Coherent 899), fixed in frequency, passes through the cell and onto a third photodiode. The frequency of this laser is locked to a saturated absorption feature in a room temperature iodine cell, and then shifted by an acousto-optic modulator to be resonant with a particular transition in the molecule (the Q(0) transition - see later). The absorption of this normalisation beam measures the number of ground state YbF molecules produced and is used to normalize the spectroscopic data with respect to the variation in the YbF yield.

The ablation laser is fired at 10Hz, and is synchronised with the 50Hz line frequency to suppress the effects of line noise in the experiment. After each ablation pulse the output from each of the three photodiodes is recorded for 8\,ms with a sample rate of 250kHz, and the transmission of each of the optical cavities is measured. The frequency of the spectroscopy laser is stepped between each shot with a typical scan consisting of 800 points covering a range of 1-3\,GHz.

To obtain high quality saturated absorption spectra, the experimental parameters need to be chosen carefully. The primary mechanism leading to the saturation of the absorption is depletion of the resonant state by optical pumping into other, non-resonant, states. Collisions of the molecules with the buffer gas repopulate the depleted state. Thus, the saturation of the absorption relies on a competition between optical pumping and collisional redistribution, and the laser intensity required for saturation depends on the buffer gas density. If the density is high the required laser intensity will be high and the saturated absorption features will be pressure broadened. If the buffer gas density is too low, the molecules will not adequately thermalize and the molecular density will drop too quickly by diffusion to the walls. We find helium densities in the range $10^{16}-10^{17}$\,cm$^{-3}$ to be most suitable for these experiments. At these densities, and for typical elastic collision cross-sections of order $10^{-14}$\,cm$^{2}$, collisions will not significantly broaden an electric dipole allowed optical transition whose natural linewidths are typically of order 10\,MHz. The diffusion time of the molecules in our cell is of order 1\,ms at these densities, and the laser intensity needed to saturate the absorption is of order 100\,mW/cm$^{2}$. The energy and repetition rate of the ablation laser is another important parameter. If these are too high there will be too much heat deposited into the cell for low temperature experiments, but if too low the density of molecules produced, and hence the degree of absorption, is too low. The parameters used here optimize the yield of YbF and raise the cell temperature by a few Kelvin.


\section{Results}
\label{sec:results}

We study the spectrum of the YbF $X {}^2\mathrm{\Sigma}^+ (v\!=\!0)$--$A {}^2\mathrm{\Pi}_{1/2} (v\!=\!0)$ transition at 552\,nm, probing the $[\mathrm{Q}_1,{}^{\mathrm{Q}}\mathrm{R}_{12}]$ and $[\mathrm{P}_1,{}^{\mathrm{P}}\mathrm{Q}_{12}]$ branch transitions, which we henceforth refer to as $^{i}$Q($N$) and $^{i}$P($N$), $N$ being the rotational quantum number in the electronic ground state. The superscript $i$ denotes the isotopologue, which we shall only include when necessary.

\begin{figure}
\begin{centering}
\includegraphics[width=0.75\textwidth]{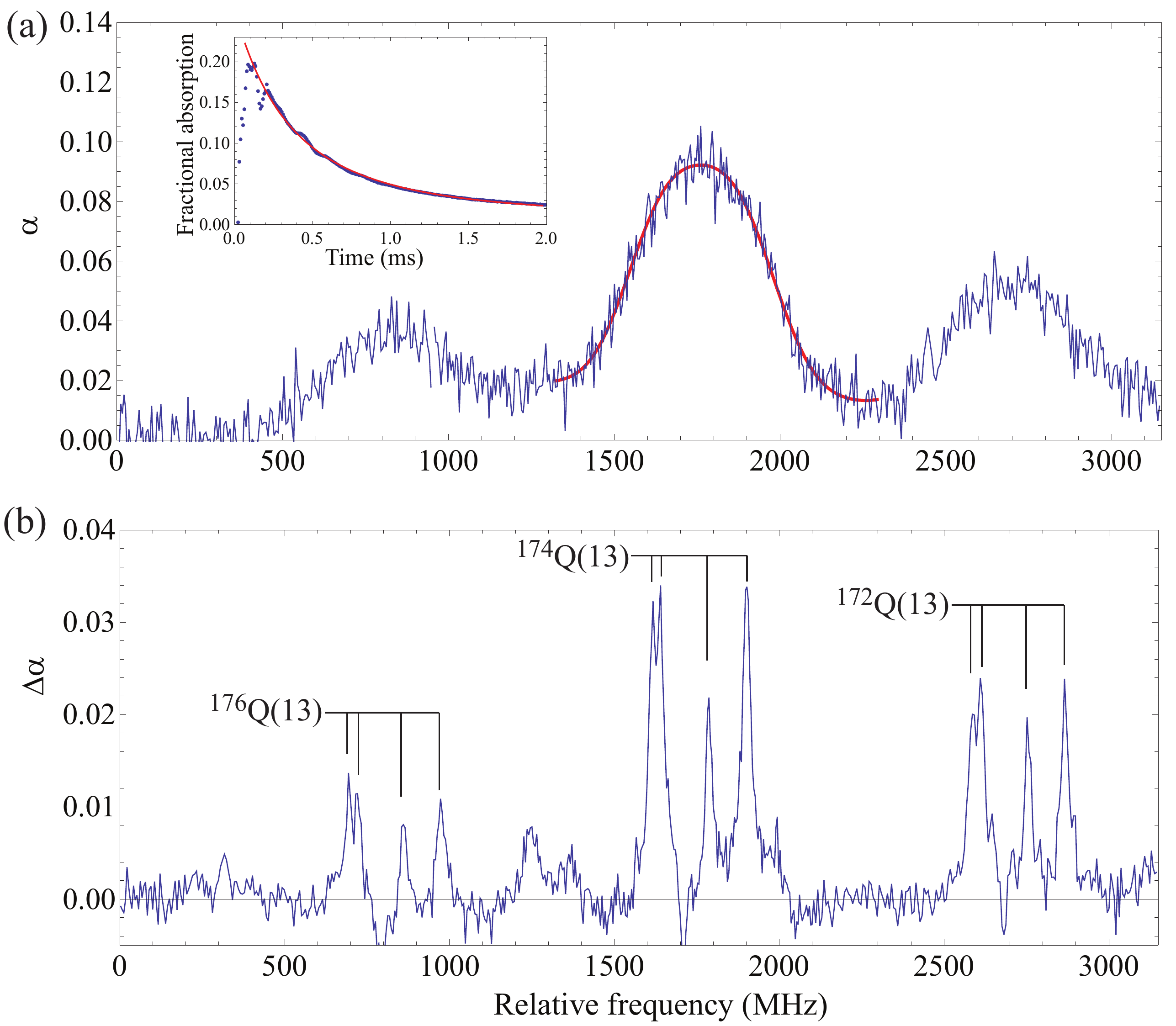}
\caption{\label{fig:dopplersatspecexample}Part of the spectrum containing the Q(13) transition of the various YbF isotopologues. (a) Doppler-broadened  absorption spectrum along with a fit to the central peak as discussed in the text. The inset shows the fractional absorption as a function of time following a single ablation shot, along with a fit (solid line) to a double exponential decay. (b) Doppler-free saturated absorption spectrum showing the hyperfine structure of the same transitions.}
\end{centering}
\end{figure}

Figure \ref{fig:dopplersatspecexample} presents some typical data, showing the region of the spectrum containing the Q(13) line of the various YbF isotopologues, obtained with a cell temperature of 88\,K and buffer gas density of $1.4 \times 10^{16}$\,cm$^{-3}$. At each laser frequency, the signal from each photodiode measures the fractional absorption as a function of time. An example absorption profile is shown in the inset of Fig.~\ref{fig:dopplersatspecexample}(a). The ablation laser is fired at time $t=0$ and there is a fast initial rise in absorption with a rise time that is limited by the detectors. Following this, a decay in the absorption signal is observed whose time constant is 0.5--1\,ms at typical operating pressures. This decay is principally due to diffusion of the molecules to the cell walls. The fractional absorption, $A$, is often as large as 50\%. In the low intensity limit, where the transition is not saturated, $A$ is related to the optical depth, $\alpha$, by $\left( 1 - A \right) = \mathrm{e}^{-\alpha}$. When the laser is on resonance with a transition from rotational state $N$, the optical depth is proportional to the density of YbF molecules, the fraction in state $N$ (which depends on the rotational temperature), and the absorption cross-section. Using a model of the absorption in the cell that includes optical pumping of the molecules and collisional redistribution amongst the rotational and vibrational states, we estimate that approximately $10^{12}$ ground state YbF molecules are produced in each ablation shot, with densities of approximately $10^{11}$\,cm$^{-3}$.

Spectra are obtained by plotting the value of $\alpha$, averaged over some fixed time window, as a function of the laser frequency. For the data shown in Fig.~\ref{fig:dopplersatspecexample} we have averaged over the time window from 1.2\,ms to 1.6\,ms. Even in these late time windows, we obtain spectra with a good signal-to-noise ratio. We divide the spectrum obtained from the reference beam by that from the normalisation beam to give a normalised absorption spectrum, which exhibits Doppler-broadened lines as shown in Fig.~\ref{fig:dopplersatspecexample}(a). Carrying out the same procedure using the probe beam yields a similar normalised absorption spectrum that contains the saturated absorption features. It is by taking the difference between these two normalised spectra that we obtain the Doppler-free saturated absorption spectrum shown in Fig.~\ref{fig:dopplersatspecexample}(b). The hyperfine structure of the even isotopologues consists of four components which are completely unresolved in the Doppler-broadened spectrum but are revealed in the Doppler-free spectrum. This structure arises from the interaction of the electron spin with the molecular rotation and with the spin of the fluorine nucleus. As we show later, the hyperfine constants can be determined by measuring the hyperfine intervals over a range of $N$ from spectra similar to the one in Fig.~\ref{fig:dopplersatspecexample}(b).

The full width at half maximum (FWHM) of the lines in the Doppler-free spectrum is typically 30\,MHz. Previous high resolution spectroscopy on a beam of YbF set an upper limit of 15\,MHz to the natural linewidth of the A state \cite{Wang96}. The corresponding values for the structurally-similar molecules, BeF, MgF, CaF and SrF, all lie in the range 7--23\,MHz \cite{Dagdigian74, Pelegrini05, Wall08}, the larger values corresponding to the shorter wavelengths as is expected for similar transition moments. It is reasonable to assume that YbF follows this trend and so has a natural linewidth of approximately 10\,MHz. The linewidth we observe is due to power broadening by the high intensity pump beam. Our choice of pump intensity yields a good compromise between the amplitude and the width of the Doppler-free lines. A lower intensity results in a poor signal-to-noise ratio, and a higher intensity leads to excessive broadening. Transit time broadening is negligible in the experiment because the molecules spend tens of microseconds diffusing through the laser beams. Zeeman broadening is at the 1\,MHz level, and so is also negligible. Imperfect collimation and imperfect alignment of the counter-propagating laser beams both lead to a residual Doppler broadening of the saturated absorption features, but using upper limits for these imperfections we find this to be negligible. Furthermore, residual Doppler broadening would be larger at higher temperatures, but we found no evidence for that. At the buffer gas densities we use, calculations using a typical elastic collision cross-section suggest that pressure broadening is not significant, and we did not observe any change in the linewidth when we varied the density up to $2\times10^{17}$\,cm$^{-3}$.

The translational temperature of the molecules can be determined from the width of the Doppler-broadened absorption lines such as the ones shown in Fig.~\ref{fig:dopplersatspecexample}(a). The unresolved hyperfine structure broadens the line and so we fit a blend of four Gaussians to the Doppler-broadened line, using a common width for each component and with the line positions fixed in the fit using the values obtained from the saturated absorption spectrum. Thus, the Doppler-free spectrum is essential for an accurate determination of the translational temperature. For the data shown in Fig.~\ref{fig:dopplersatspecexample}, the translational temperature obtained from this procedure is $91\pm8$\,K, consistent with the cell temperature of 88\,K. We typically find that in the 1.2--1.6\,ms time window, the translational temperature is 78$\pm$8\,K when the cell temperature and gas density are 80\,K and $2.4 \times 10^{16}$\,cm$^{-3}$, indicating efficient thermalization of the molecules under these conditions. When the temperature and gas density are 10\,K and $4.6 \times 10^{16}$\,cm$^{-3}$ we measure a translational temperature of 17$\pm$1\,K suggesting that the molecules take longer to reach the base temperature in this case. Our analysis of how the translational temperature evolves suggests that the helium gas is heated by the ablation process and that it is the re-thermalization of the helium with the cell walls that limits the cooling process, rather than the thermalization of the molecules with the helium. When the helium is colder it moves more slowly and takes longer to thermalize with the walls.

\begin{figure}
\begin{centering}
\includegraphics[width=0.9\textwidth]{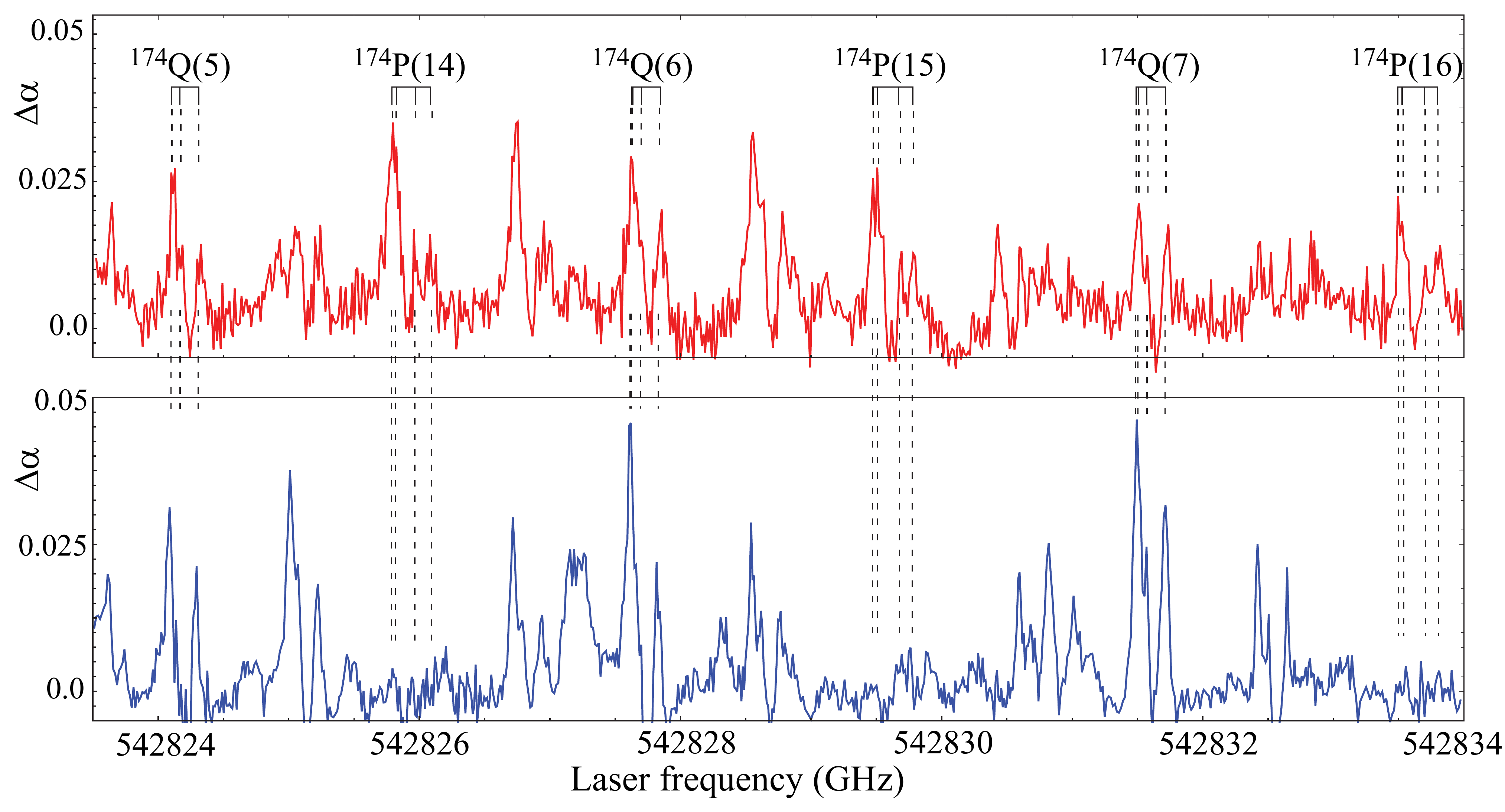}
\caption{\label{fig:hotandcoldsatabs}Saturated absorption spectra recorded with a buffer gas cell temperature of 80K (upper trace) and 10K (lower trace).  At low temperature, transitions from the higher rotational states (the P lines) have vanished. Spectral lines that are not labelled are due to the other isotopologues of YbF.}
\end{centering}
\end{figure}

A key benefit of our technique is the ability to control the temperature of the molecules simply by changing the temperature of the cell, whilst maintaining the narrow, Doppler-free spectral lines. This allows us to tailor the rotational population to optimise the lines of interest in a given spectral region. At very low temperatures the population is confined to the lowest rotational states, providing clear, uncluttered spectra. At higher temperatures population is distributed over a larger number of states and lines corresponding to higher rotational states can be probed. Fig.~\ref{fig:hotandcoldsatabs} shows an example where we have scanned over the spectral region containing the Q-branch lines from $N=5$ to $N=7$, at two different cell temperatures, 80\,K and 10\,K. The same spectral region contains P-branch transitions from $N=14$ to $N=16$. These are visible at the higher temperature, considerably complicating the spectrum, but almost vanish at the lower temperature because of the reduced population in the higher-lying rotational states. This temperature control makes it far easier to identify the two sets of lines and to measure the hyperfine intervals over a large range of $N$.

The relative intensities of the lines in the spectrum can be used to determine the rotational temperature of the molecules. The intensities follow a Boltzmann distribution slightly modified by the rotational state dependence of the transition matrix elements. Applying this procedure to the spectrum observed in the 1.2--1.6\,ms time window, we measure rotational temperatures of 100$\pm$7\,K and 42$\pm$5\,K for cell temperatures of 80\,K and 10\,K and densities of $2.4 \times 10^{16}$\,cm$^{-3}$ and $4.6 \times 10^{16}$\,cm$^{-3}$ respectively. This shows that the rotational temperature is slower to thermalize than the translational temperature at these buffer gas densities. We will report the measurement of the collision cross-sections elsewhere, along with the details of how we measure the YbF number density and a more thorough investigation of the formation, diffusion, cooling and loss of molecules as a function of the experimental parameters.

\begin{figure}
\begin{centering}
\includegraphics[width=0.75\textwidth]{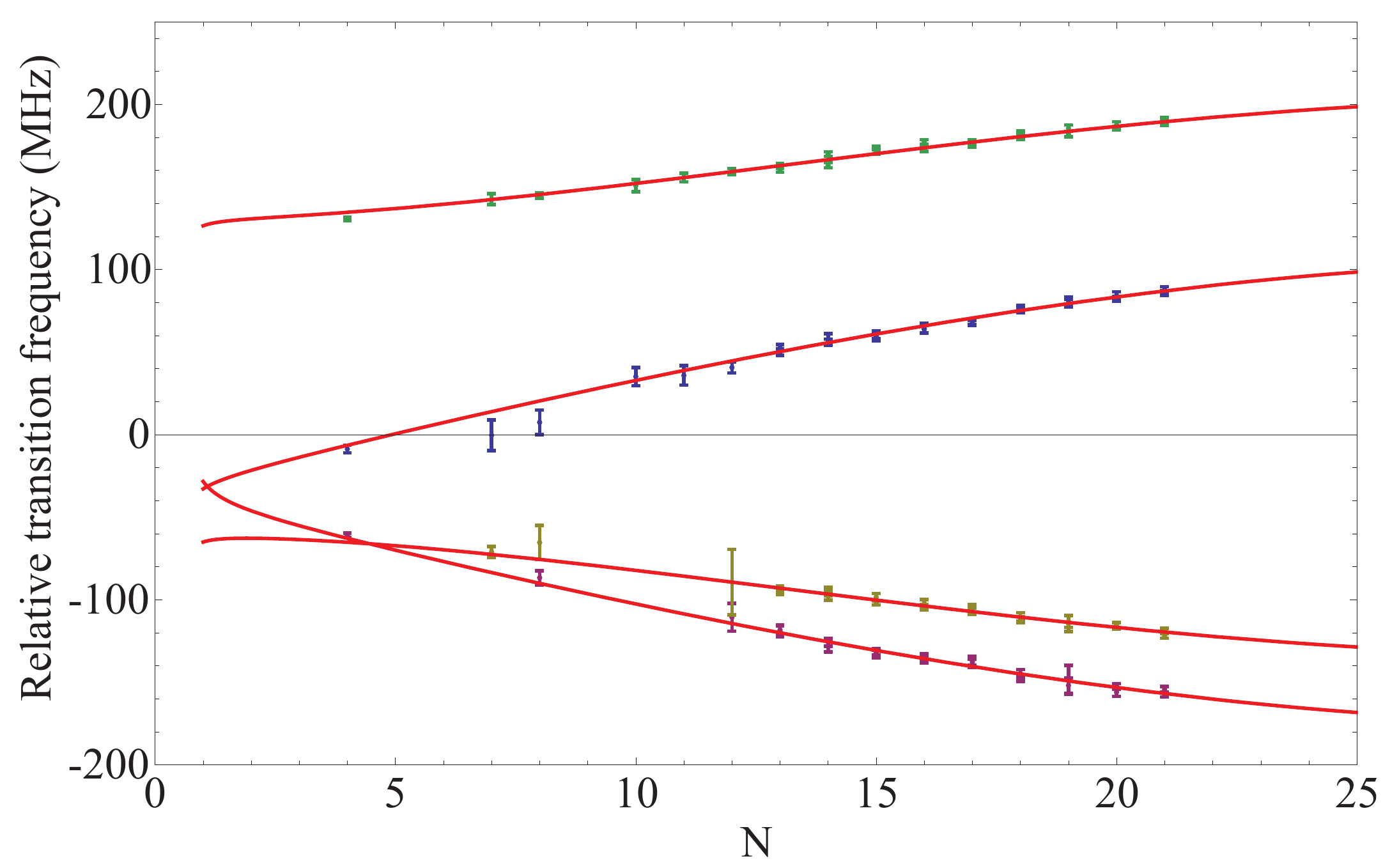}
\caption{\label{fig:fishfit}Measured frequency intervals of the four hyperfine components of $[\mathrm{Q}_1,{}^{\mathrm{Q}}\mathrm{R}_{12}]$ branch transitions as a function of rotational state. At each $N$ we have plotted the frequency of each hyperfine component minus the mean frequency of all four. Fitting these data to the eigenvalues of the effective hyperfine Hamiltonian \cite{Sauer96} determines the hyperfine parameters and gives the best fit lines shown.}
\end{centering}
\end{figure}

To demonstrate further the usefulness of this technique for precise optical spectroscopy at high resolution, we have measured the separation of the four hyperfine components of the $^{174}$YbF $[\mathrm{Q}_1,{}^{\mathrm{Q}}\mathrm{R}_{12}]$ branch transitions originating from rotational states between $N$=4 and $N$=12 using a cell temperature of 10\,K, and between $N$=13 and $N$=21 with a cell temperature of 80\,K. From these measured intervals, and the effective hyperfine Hamiltonian given in \cite{Sauer96}, $H_{\rm eff}=\gamma {\bf S}.{\bf N} + b {\bf I}.{\bf S} + c I_{z}S_{z}$, we determine the spin--rotation coupling parameter $\gamma = \gamma_0 + \gamma_1 N \left( N+1 \right)$, and the hyperfine interaction parameters $b$ and $c$. The measured intervals, along with the best fit to this hyperfine Hamiltonian are shown in Fig.\,\ref{fig:fishfit}. Our best-fit parameters are $\gamma_{0}=-13.52 \pm 0.23$\,MHz, $\gamma_{1}=4.72 \pm 0.64$\,kHz, $b=139.8 \pm 3.9$\,MHz and $c=86 \pm 12$\,MHz. These results are in good agreement with those of similar precision obtained by laser spectroscopy of an effusive beam \cite{Sauer95}, and with the more precise values obtained using the laser-rf double resonance technique reported in \cite{Sauer96}.


\section{Summary and outlook}
\label{sec:conclusions}

We have shown how to obtain high resolution optical spectra of molecular species that are difficult to produce in the gas-phase using a simple and versatile apparatus that cools the molecules and allows the temperature and pressure to be controlled over a very wide range. Control of the rotational temperature is very helpful in identifying lines in a congested spectrum and making measurements over a broad range of rotational states. The Doppler-free spectra are also valuable for accurate measurements of the translational temperature of the molecules. Key to the success of the technique is the ability to produce optically thick samples of radicals by laser ablation into the cold buffer gas. Using the same techniques, molecular densities similar to ours have been reported for several other species e.g. \cite{Weinstein(3)98, Maussang05, Lu09}, so we expect that laser ablation of suitable targets will allow a very wide range of other molecules to be produced with optical densities sufficient for saturated absorption spectroscopy. Doppler-free polarization spectroscopy, whose sensitivity is often superior to saturated absorption spectroscopy \cite{Wieman76}, could also be applied to cold molecules produced this way. The use of a cryocooler would further simplify these experiments by eliminating the need to handle cryogenic liquids and providing very fast cool-down times.


\ack We are grateful to Jon Dyne, Steve Maine, Valerijus Gerulis and Robert Finnis for their expert technical assistance. This work was supported by the EPSRC and the STFC.

\section*{References}

\bibliography{buffer_sat_spec}

\end{document}